\documentclass[aps,pra,epsfig,twocolumn]{revtex4}
\usepackage{graphicx}% Include figure files
\usepackage{dcolumn} % Align table columns on decimal point
\usepackage{bm}
\usepackage{amsmath}
\begin{document}

\draft

\title{Quantum Communication Through Spin Chain Dynamics: An Introductory Overview}

\author{Sougato Bose}

\affiliation{Department of Physics and Astronomy, University
College London, Gower St., London WC1E 6BT, UK}

\date{\today}

\begin{abstract}
We present an introductory overview of the use of spin chains as
quantum wires, which has recently developed into a topic of lively
interest. The principal motivation is in connecting quantum
registers without resorting to optics. A spin chain is a permanently
coupled 1D system of spins. When one places a quantum state on one
end of it, the state will be dynamically transmitted to the other
end with some efficiency if the spins are coupled by an exchange
interaction. No external modulations or measurements on the body of
the chain, except perhaps at the very ends, is required for this
purpose. For the simplest (uniformly coupled) chain and the simplest
encoding (single qubit encoding), however, dispersion reduces the
quality of transfer. We present a variety of alternatives proposed
by various groups to achieve perfect quantum state transfer through
spin chains. We conclude with a brief discussion of the various
directions in which the topic is developing.
\end{abstract}
 \maketitle

%\begin{multicols}{2}
\section{Introduction}
Quantum communication is the act of transferring a quantum state
from one place to another. By far its most well known application is
quantum key distribution through which a secret random key can be
established between distant parties with its security guaranteed by
quantum mechanics \cite{BB84,E91}. In quantum key distribution, a
quantum state prepared by one party only needs to be measured by
another party at a distance. For this purpose, photons are very well
suited as they easily travel long distances through optical fibres
or empty space and can be readily measured by a receiving party.
Increasingly, however, the pivotal importance of quantum
communication in a different area of quantum information processing,
namely quantum computation, is becoming clear. It is becoming
important for connecting up distinct quantum processors or registers
to make a powerful quantum computer \cite{wineland,duan,kane03}. For
this application, it is not only important to transfer a quantum
state between locations but also to map it from/to the elements of
the quantum register sending/receiving it. This necessitates simple
exchange of quantum information between the elements of a quantum
computer and the entities carrying the information between the
computers. Moreover the transfer is needed only over short distances
separating distinct registers. For such short distance applications,
where mapping of the transferred quantum state to the elements of a
register is also important, it is very useful to have alternatives
to photons \cite{wineland,kane03}. This review article will be based
on one such alternative, where the quantum state transfer is
accomplished purely through the natural dynamical evolution of a
permanently coupled chain of quantum systems, which has recently
drawn considerable attention
\cite{bose02,subrahmanyam04,amico04,christandl04,petrosyan,yung03,osborne04,eisert04prl,eisert04,haselgrove04,albanese04,vittorio04,burgarth04,li04,shi04,
plenio04,yung05,korepin04,vittorio05,romito,burgarth05,paternostro,bayat,li05,christandl05,dechiara05,vaucher,karbach,yang05a,twamley,fitzsimons06,greentree,
qian05,hamieh,wojcik05,paternostro05,bruder05,hadley05,burgarth-giovannetti,kay06,perales05,yang05b,vittorio-fazio06,dechiara-fractal,wojcik06,bruder06,zong-chen,
kay2006a,depascualle06,guo06,boness06,avellino06,keating06,burgarth-giovannetti-bose,yung06a,burgarth-bose06,hartmann05,sanpera}.
We start by highlighting the importance of short distance quantum
communications in quantum computation and why non-photonic
alternatives are important to pursue. At this point it is worth
mentioning that dynamics
    is certainly not the only way to transfer a quantum state through a chain of permanently coupled quantum systems. Stationary
    states (such as thermal and ground states) of many-body systems have
    been studied from the quantum information theoretic viewpoint for quite a while
    (see, for example,
    Refs.\cite{arnesen00,osterloh-amico,osborne-nielsen,vidal-rico,revmod})
    and recently it was realized that certain classes of such chains accomplish the perfect
    transfer of a quantum state through measurements \cite{verstraete-cirac}, while certain other classes can accomplish
    the same even without measurements \cite{campos-venuti,rigolin}. However, in this review, I will concentrate
     only on the literature on the use of spin chain dynamics for quantum communications.

\begin{figure}
\begin{center}
\includegraphics[width=3.5in, clip]{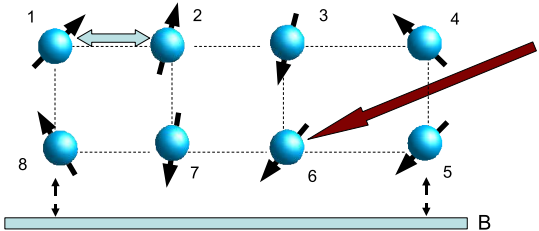}
 \caption{The figure shows a typical quantum computer. The spheres with arrows denote qubits such as spin-1/2 systems.
 Typically operations on individual qubits are
required, and accomplished by local
 fields on individual qubits, as shown, for example, by the long arrow controlling qubit 6. Moreover, any pair of qubits are required to interact
 with each other. This can take place through a common bus. For example, qubits 5 and 8 are shown to be interacting through the common bus B. Naturally,
 the number of qubits which can be accommodated on a single bus is limited, and this restricts the size of feasible computers. An alternate strategy,
 in computers without a common bus mode is to make qubits interact directly. In those cases, if one intends to make joint operations
 between distant
 qubits, then their states have to be transferred to neighbouring qubits, such as qubits 1 and 2, and then these qubits will be made to
interact, as shown. For efficient functioning of the computer, the
time-scale of the transfer must be restricted, which again limits
the size of the computer.}
\label{qc}
\end{center}
\end{figure}

  \subsection{Quantum Communication for Connecting Quantum Registers/Computers:}   Quantum Computers, when
realized, hold the promise of speeding up
 the solution of certain problems perceived as difficult on a classical
 computer \cite{D-J,shor,grover,NC}. They also hold the promise of
 enabling controlled simulations of the behaviour of
 complex quantum systems \cite{feynman,NC}. The typical quantum computer is regarded
 as a collection of quantum two state systems (or ``qubits") on which arbitrary unitary operations
 can be performed, as for example shown in Fig.\ref{qc}. The power of a quantum computer increases with the number of
qubits. However, there are several
 fundamental obstacles to increasing the number of qubits in a quantum computer arbitrarily.
 For example,
  the computer may be based on a common bus through which the qubits interact \cite{cirac-zoller,imamoglu,cleland}, as depicted
  in Fig.\ref{qc}. Then
 there is a physical limitation on the number of qubits which can be linked by the same bus.
 Alternatively, if the computer is based on direct interactions between
 qubits \cite{loss-divincenzo,kane,barnes}, as also shown in
 Fig.\ref{qc},
 then either the
qubits have to be moved close enough to interact, or the states
they bear have to be transferred to qubits which are already
within the range of each other's interaction. Thus the size of an
individual quantum computer will be limited by the need to
maximize efficiency. One way to get around the above problems is
to envisage a quantum computer composed of a number of quantum
registers connected to each other by quantum communication
channels (which could effectively be physical shuttling of qubits
\cite{wineland,kane03}). Most operations would take place between
qubits of the same register through, say, a common bus mode, or
direct interactions together with very fast qubit transfers within
the register. Occasionally, the quantum channels would be used to
transfer qubits from one register to another, and thereby enable
quantum gates between qubits of different registers. In contrast
to the quantum channels required for quantum key distribution,
these channels inside a quantum computer could be very short and
should allow significant pre and post processing of the quantum
state communicated through the channel in the transmitting and
receiving registers.

 Even if we are able to scale up quantum computers by
some technology which does not require internal communication
channels, such channels will still be necessary to hook up
distinct quantum computers. For reasons of compactness, mobility
and cost, we might just prefer to have small sized quantum
computers. However at times one may need to tackle very complex
problems for which the power of a single quantum computer will not
suffice. It will then become very important to combine the
processing powers of distinct quantum computers to obtain a
computer with a greater processing power. Again, in contrast to
the photonic channels currently being used for quantum key
distribution, these need not be long (the computers could be
sitting next to each other). Moreover, as the channel connects
quantum computers, some encoding and decoding of the state should
easily be possible inside the quantum computers. The transmitted
state (after decoding, if applicable) should be transferable to a
qubit or a group of qubits of the quantum computer that receives
it.

\begin{figure}
\begin{center}
\includegraphics[width=3.5in, clip]{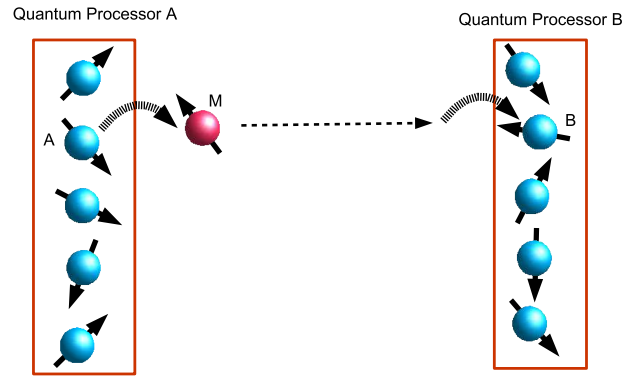}
 \caption{The usual approach to connecting quantum processors. Quantum information stored in the stationary (blue) qubits
 of one register is mapped to a mobile (red) qubit. This qubit flies to the other register, where the information stored in it is
 mapped on to the stationary (blue) qubits of that register.}
\label{transa}
\end{center}
\end{figure}

\subsection{An unmodulated chain of quantum systems as a communication channel between quantum computers}
\begin{figure}
\begin{center}
\includegraphics[width=3.5in, clip]{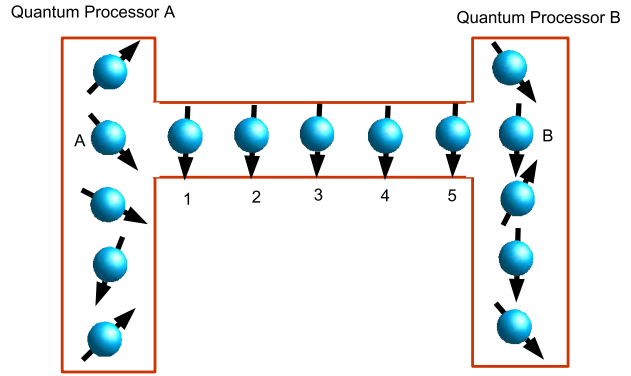}
 \caption{This figure depicts the possibility of transporting quantum information from one quantum processor to another through a line of
 stationary qubits.}
\label{transb}
\end{center}
\end{figure}

The usual approach envisaged for connecting quantum computers (or
processors) is to first map the state to be transmitted from the
qubits of one processor to a {\em flying} or {\em mobile} qubit.
This flying qubit then traverses through a channel and reaches a
second processor, where its state is mapped on to the qubits of that
processor. The usual approach is depicted in Fig.\ref{transa}.
However, depending on the physical nature of the qubits of the
processors, this approach could involve either (a) interfacing
between different physical systems such as stationary spins and
photons \cite{yablonovich} or stationary and mobile spins
\cite{costa2} or (b) physically moving a quantum system and
subsequently bringing it to a halt elsewhere such as shuttling ions
\cite{wineland} or electrons \cite{kane03}. All the above can be
complicated in many respects. So one can ask the question: is it
possible to transfer quantum information from place to place {\em
using only stationary qubits}? The first idea that comes to mind is
to have a chain of qubits as shown in Fig.\ref{transb}, and swap a quantum state perfectly in
succession from one qubit to the next.
A quantum state encoded on a qubit at one end of the line can be
transported perfectly through a series of swaps to the qubit at the
opposite end of the line. For example, in Fig.\ref{transb}, the
strategy is to swap state in the following order: $A\rightarrow
1,1\rightarrow 2,2\rightarrow 3,3\rightarrow 4,4\rightarrow
5,5\rightarrow B$. This kind of data-bus, called a {\em swapping
channel}, has been discussed, for example, in Ref.\cite{chuang}.
However, such a data-bus requires the ability to modulate the
strength or nature of interactions between pairs of adjacent qubits
(such as $1$ and $2$ or $3$ and $4$) in time. Typically, this would
require control fields on the wire varying over the scale of the
spacing between the qubits. If so much control is available on a
chain of qubits, then why not use the chain as a quantum
computer? It will then be a gross under-utilization to use it merely
as a data-bus. Moreover, the requirement of so much control for the
transfer of a quantum state naturally implies that such a protocol
is also very susceptible to errors in these controls. For example,
there are 6 pair-wise interactions to be switched on and off in
succession for transmission of the state of qubit $A$ to that of
qubit $B$ in Fig.\ref{transa}, and errors would accumulate in each
of these steps. Thus the question arises as to whether we can
utilize systems with much lower controls for connecting quantum
registers. For example, if the interactions between qubits in a
chain are permanent and uncontrollable (always on and constant in
strength), and we are not allowed to apply any control fields to the
qubits, could the chain still act as a quantum data-bus? The
validity of the above possibility will enable us to use such a qubit
chain as a data-bus in the true sense of word. This is because in
the normal everyday use of the word ``data-bus", such as to denote a
cable connecting two computers, we do not envisage controlling every
individual part of the cable and we mostly let the information flow
through it in its own natural way. A qubit chain in which
inter-qubit interactions are permanent, is an example of a {\em spin
chain}, which we introduce below.

\subsection{Spin chains and the exchange interaction}
\begin{figure}
\begin{center}
\includegraphics[width=3.5in, clip]{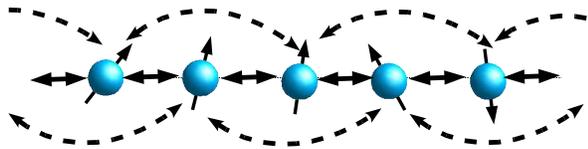}
 \caption{The figure shows a spin chain: a system of spins perpetually coupled to each other with
 an interaction strength which generally decreases with distance. The double arrowed lines depict interactions with
 the dotted line denoting a weaker strength than the solid line.}
\label{spinchain}
\end{center}
\end{figure}

As the reader will know, in quantum mechanics, spins are systems
endowed with tiny quantized magnetic moments. Bulk materials often
have a large collection of spins permanently coupled to each other.
The mutual interactions of a of these spins makes them prefer
alignment or anti-alignment with respect to each other, resulting in
diverse phenomena such as ferromagnetism and anti-ferromagnetism. A
spin chain models a large class of such materials in which the spins
are arranged in a one dimensional lattice and permanently coupled to
each other, usually with an interaction strength decreasing with
distance (as shown in Fig.\ref{spinchain}). A common form of the
Hamiltonian for the interaction between the $i$th and the $j$th spin
is written as
\begin{equation}
{\bf H}_{ij}=J_{ij}~\vec{S}_i.\vec{S}_{j}, \label{heis}
\end{equation}
where $\vec{S}_i.\vec{S}_{j}\equiv S^x_iS^x_j+S^y_iS^y_j+S^z_iS^z_j$
and $S^x_i,S^y_i,S^z_i$ are the operators for the component of the
$i$th spin along the $x,y$ and $z$ directions respectively. In
particular, when all the spins are spin-1/2 systems, $S^x,S^y$ and
$S^z$ stand for the familiar Pauli matrices $\sigma^x,\sigma^y$ and
$\sigma^z$. A Hamiltonian of the above form is termed as an {\em
exchange interaction} as it can arise in from the pure exchange
electrons between neighbouring ions in a metal. It is also called
the {\em Heisenberg interaction} after its inventor. In particular,
the specific Hamiltonian we have written above is called the {\em
isotropic} exchange interaction. In this paper we will also
encounter a variant of the above interaction
\begin{equation}
{\bf H}_{ij}^{XY}=J_{ij}~(S^x_iS^x_j+S^y_iS^y_j),
\end{equation}
which is called the XY interaction. We will be primarily concerned
with chains of spin-1/2 systems in this article. Not only do examples of such
systems exist in nature \cite{hammar}, but also can be fabricated in
systems of any kind of qubits \cite{romito,bruder05,bruder06}, as qubits are
isomorphic to spin-1/2 systems. If one indeed fabricates artificial
systems, why would one fabricate a spin chain
 {\em i.e.,} a system with permanent couplings rather
than a system where such couplings are also switchable? The obvious answer is that such a system should have a much
lower complexity of fabrication because they do not require the an
attached mechanism (such as electrodes) varying
over the scale of the separation of the qubits to
modulate their interactions.

\subsection{Quality check for a qubit array} We now provide a couple of other motivations
for studying quantum communications through a spin chain, quite
distinct to the simple aim of connecting quantum registers. Suppose
a linear array of qubits has been fabricated to function as a
quantum computer. One then needs to check whether each qubit behaves
as a bonafide qubit: namely they are able to remain in a quantum
superposition of two distinct states for a time longer than that
needed for running an entire quantum algorithm in the computer. Such
superpositions may be destroyed by the interaction of a qubit with
its environment, a process known as decoherence. The most
straightforward strategy will be to probe each qubit individually by
first switching off a qubit's interactions with adjacent qubits,
initializing it in a known state and then applying a set of known
unitary operations to it and measuring its state at the end. This
strategy is very time consuming for a long array. An easier way is
just to test the ability of the array to behave as quantum
communication channel. One has to place a known quantum state on the
qubit at one end of an array, switch on the interactions of each
qubit with its immediate neighbours (so that the qubit array is now
isomorphic to a spin chain with nearest neighbour interactions) and
probe how well the state is retrieved from the other end after a
specified interval of time. This requires initialization and
measurements on only the two qubits at the ends of the array. As the
quantum state has to pass through {\em all} the intermediate qubits in
order to be transmitted from one end of a linear array to another,
the ability of each qubit to behave as a bonafide qubit is
automatically tested. If there are a few faulty qubits in the array
(faulty means that they decohere significantly in the time-scale of
the experiment), then the quality of the state transmission will be
much lower than that expected for a fault-less array. Thus quantum
communication through an array of qubits can enable a quality test
of the entire array by just probing two of the qubits.

\subsection{Quantum response to a quantum impulse}
Typically, one characterizes the behaviour of complex systems by how
it responds to external stimuli. Usually a classical field is applied to
some part of the system and the resulting change in the value of
some variable at a different part of the system is determined. For
example, one can characterize magnetic systems by their magnetic
susceptibility, which can determine how a magnetic field applied to
one part of the system affects the magnetization in a different part
of the system. Knowledge of this response enables the design of
components which may use the magnetic system under consideration. In
recent years, our ability to manipulate and measure single quantum
systems have improved. It thus makes sense to speak about a fully
quantum counterpart of the usually studied stimulus and response. We
provide a quantum stimulus by placing a quantum state at one part of
a complex system and study its quantum response: the quantum state
produced at another part of the system after some time as a {\em result}
of the placement. Quantum communication through a spin chain can be
regarded as a specific example of the above study, where one puts a
quantum state at one end of a specific complex system (namely the
spin chain) and looks at how well the spin at the other end of the
chain resembles that state as a function of time.

\begin{figure}
\begin{center}
\includegraphics[width=3.5in, clip]{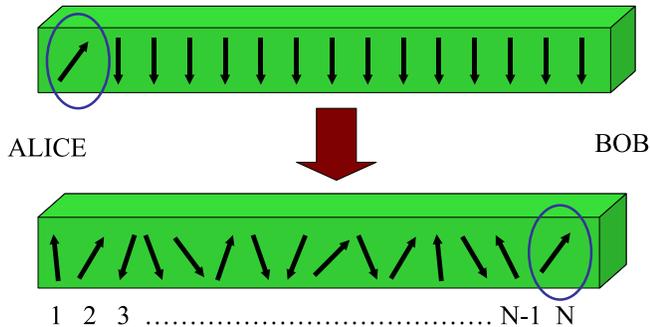}
 \caption{The simplest spin chain communication protocol. A spin-1/2 ferromagnetic
 spin chain with all spins facing down is the quantum channel. Alice simply places a quantum state at one end of the
 chain and Bob simply picks up a close approximation of this state from his end of the spin chain after waiting a while.}
\label{simplest-scheme}
\end{center}
\end{figure}

\section{The simplest spin chain quantum communication protocol}
We will now present the simplest spin chain communication protocol
that one can envisage and analyse its performance. The simplest
possible spin chain is one composed of the lowest dimensional spin
systems, namely spin-1/2 systems or qubits (so that the operators
$S^x,S^y$ and $S^z$ can be replaced by the familiar Pauli matrices
$\sigma^x,\sigma^y$ and $\sigma^z$ ). We also want to start our
protocol with the spin chain initialized in a very simple state,
such as one in which all spins are in the same pure state, say
$|0\rangle$. We will have to choose the couplings $J_{ij}$ in
Eq.(\ref{heis}) in such a manner that initialization of the spin
chain to such a state is easy. Accordingly, we choose $J_{ij}<0$,
which means that the spin chain describes a {\em ferromagnet}. The
ground state of a ferromagnet in a magnetic field, however weak, is
one where all spins point in the direction of the field. For example,
all spins could be pointing down. In the simplest quantum
communication protocol, Alice places an arbitrary quantum state at
one end of the spin chain in such a ``all down" state.
This is depicted in the upper part of
Fig.\ref{simplest-scheme}, where Alice has placed an arbitrary state on the first spin of the chain, while all the other
spins are still in the down state.  Due to the natural evolution of the chain
this state both disperses and propagates along the chain. As a
result of this evolution, the state of the spin at Bob's end of the
chain will vary with time. Bob now chooses an optimal moment of time
in as long an interval that he can afford to wait to receive Alice's
state. This moment of time is carefully chosen so that the state of
the spin at Bob's end of the chain is as close as possible to the
one that Alice intended to transmit. At this optimum time, Bob
simply picks up the spin at his end of the chain, to conclude the
communication protocol.

\subsection{Fidelity as a figure of merit} In order to judge how
well a quantum state is transferred by a spin chain one has to use a
figure of merit. Suppose the state that Alice places into the spin
at her end of the chain is depicted by $|\psi_{in}\rangle$ and the
state of the spin that Bob picks up at the optimum time $t_0$ is
depicted in general by the density operator $\rho_{out}(t_0)$ (the
output state is depicted by a density operator to allow for the
possibility for it to be a mixed state). Then a measure of the
quality of the transfer is defined by the fidelity
\begin{equation}
F=\langle\psi_{in}| \rho_{out}(t_0)|\psi_{in}\rangle,
\end{equation}
which is always between $0$ and $1$, with higher value meaning
better transfer (it is unity for perfect transfer). A fidelity of
$2/3$ can already be obtained if Alice simply measured her state,
communicated the results classically to Bob and Bob simply
reconstructed the state from this data. Thus $F$ needs to be greater
than $2/3$ in our spin chain quantum communication scheme to be
better than straightforward classical communication. Later, we will
show that the use of another figure of merit, namely the amount of
entanglement that can be transmitted by a spin chain, aids in the
the justification that spin chains of arbitrary length are
interesting as quantum communication channels.

\subsection{General formula for fidelity for arbitrary graphs}
I will first present the scheme in a general setting for arbitrary
graphs of spins with ferromagnetic Heisenberg interactions and later
proceed to the special cases motivated by realizability. Say there
are $N$ spins in the graph and these are numbered $1,2,...,N$. The
Hamiltonian is given by
\begin{equation}
{\bf H}_{\bf G}=-\sum_{<i,j>}
J_{ij}~\vec{\sigma}^i.\vec{\sigma}^{j}-\sum_{i=1}^N B_i\sigma_z^i.
\label{ham}
\end{equation}
$\vec{\sigma}^i=(\sigma_x^i, \sigma_y^i, \sigma_z^i)$ in which
$\sigma_{x/y/z}^i$ are the Pauli matrices for the $i$th spin,
$B_i>0$ are {\em static} magnetic fields and $J_{ij}>0$ are coupling
strengths, and $<i,j>$ represents pairs of spins. ${\bf H}_{\bf G}$
describes an arbitrary ferromagnet with isotropic Heisenberg
interactions. As mentioned above we initialize the graph in its
ground state $|{\bf 0}\rangle=|000...0\rangle$ where $|0\rangle$
denotes the spin down state (i.e., spin aligned along $-z$
direction) of a spin. This can be accomplished easily for a
ferromagnetic system by cooling. We will set the ground state energy
$E_0=0$ ({\em i.e.,} redefine ${\bf H}_{\bf G}$ as $E_0+{\bf H}_{\bf
G}$) for the rest of this paper. We also introduce the class of
states $|{\bf j}\rangle=|00...010....0\rangle$ (where ${\bf j}={\bf
1},{\bf 2},..{\bf s},..{\bf r},..,{\bf N}$) in which the spin at the
$j$th site has been flipped to the $|1\rangle$ state. We now assume
that the state sender Alice is located closest to the $s$th ({\em
sender}) spin and the state receiver Bob is located closest to the
$r$th ({\em receiver}) spin. All the other spins will be called {\em
channel spins}. As mentioned before, to start the protocol, Alice
simply places the state she wants to transmit to Bob on the $s$th
spin at time $t=0$. Let this state be $|\psi_{in}\rangle=
\cos{(\theta/2)}|0\rangle+e^{i\phi}\sin{(\theta/2)}|1\rangle$. We
can then describe the state of the whole chain at this instant (time
$t=0$) as
\begin{equation}
|\Psi(0)\rangle=\cos{\frac{\theta}{2}}|{\bf
0}\rangle+e^{i\phi}\sin{\frac{\theta}{2}}|{\bf s}\rangle.
\end{equation}
Bob now waits for a specific time till the initial state
$|\Psi(0)\rangle$ evolves to a final state which is as close as
possible to $\cos{\frac{\theta}{2}}|{\bf
0}\rangle+e^{i\phi}\sin{\frac{\theta}{2}}|{\bf
   r}\rangle$. As $[{\bf H}_{\bf G},\sum_{i=1}^N \sigma_z^i]=0$, the state $|{\bf
s}\rangle$ only evolves to states $|{\bf j}\rangle$ and the
evolution of the spin-graph (with $\hbar=1$) is
\begin{equation}
|\Psi(t)\rangle=\cos{\frac{\theta}{2}}|{\bf
0}\rangle+e^{i\phi}\sin{\frac{\theta}{2}}\sum_{{\bf j}={\bf 1}}^{\bf
N} \langle {\bf j}|e^{-i{\bf H}_{\bf G}t}|{\bf s}\rangle|{\bf
j}\rangle.
\end{equation}
The state of the $r$th spin will, in general, be a mixed state, and
can be obtained by tracing off the states of all other spins from
$|\Psi(t)\rangle$. Undergraduate readers can familiarize themselves
with this tracing procedure called {\em partial tracing} from
quantum information textbooks \cite{NC}, but essentially the density
operator $\rho_{out}$ of the output state is obtained by
${\mbox{Tr}}_{1,2,...,N-1}(|\Psi(t)\rangle\langle \Psi(t)|)$, where
${\mbox{Tr}}_{1,2,...,N-1}$ means tracing over the states of the
systems $1$ to $N-1$. This evolves with time as
\begin{equation}
\rho_{out}(t)=P(t)|\psi_{out}(t)\rangle\langle\psi_{out}(t)|+(1-P(t))|0\rangle\langle
0|, \label{out}
\end{equation}
with
\begin{equation}
|\psi_{out}(t)\rangle =
\frac{1}{\sqrt{P(t)}}(\cos{\frac{\theta}{2}}|
0\rangle+e^{i\phi}\sin{\frac{\theta}{2}}
f_{s,r}(t)|1\rangle),\label{out3}
\end{equation}
where
$P(t)=\cos^2{\frac{\theta}{2}}+\sin^2{\frac{\theta}{2}}|f_{r,s}(t)|^2$
and $f_{r,s}(t)=\langle {\bf r}|\exp{\{-i{\bf H}_{\bf G}t\}}|{\bf
s}\rangle$. Note that $f_{r,s}(t)$ is just the transition amplitude
of an excitation (the $|1\rangle$ state) from the $s$th to the $r$th
site of a graph of $N$ spins.

     Now suppose it is decided that Bob will pick up the
$r$th spin (and hence complete the communication protocol) at a
predetermined time $t=t_0$. The fidelity of quantum communication
through the channel averaged over all pure input states
$|\psi_{in}\rangle$ in the Bloch-sphere ($(1/4\pi)\int
\langle\psi_{in}| \rho_{out}(t_0)|\psi_{in}\rangle d\Omega$) is then
\begin{equation}
F=\frac{|f_{r,s}(t_0)|\cos{\gamma}}{3}+\frac{|f_{r,s}(t_0)|^2}{6}+\frac{1}{2},
\label{fid}
\end{equation}
where $\gamma=\arg\{f_{r,s}(t_0)\}$. To maximize the above average
fidelity, we must choose the magnetic fields $B_i$ such that
$\gamma$  is a multiple of $2\pi$. Assuming this special choice of
magnetic field value (which can always be made for any given $t_0$)
to be a part of our protocol, we can simply replace $f_{r,s}(t_0)$
by $|f_{N,1}(t_0)|$ in Eq.(\ref{out3}).

\begin{figure}
\begin{center}
\includegraphics[width=3.3in, clip]{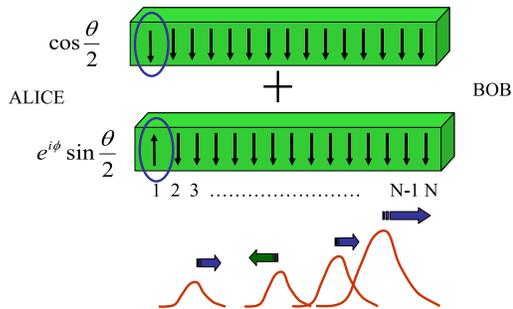}
 \caption{The figure shows how quantum information is transmitted down a spin chain in the simplest protocol.
 The chain goes to a superposition of its ground state and a time evolving state. The time evolving state transmits
 a spin flip as a series of wave-packets which travels towards Bob while dispersing at the same time.}
\label{spinwaves}
\end{center}
\end{figure}

\subsection{Specifics for an open chain with nearest neighbour interactions}
As we intend to use a graph of interacting spins as a channel, we
will use the most natural geometry for a channel, which is a linear
open ended chain (Fig.\ref{simplest-scheme}) with the sender (Alice)
and the receiver (Bob) at opposite ends. To use an analytically
solvable Hamiltonian ${\bf H}_{\bf L}$ we assume
$J_{ij}=(J/2)\delta_{i+1,j}$ (nearest neighbor interactions of equal
strength) and $B_i=B$ (uniform magnetic field) for all $i$ and $j$
in Eq.(\ref{ham}) for ${\bf H}_{\bf G}$. The eigenstates of ${\bf
H}_{\bf L}$, relevant to our problem are
\begin{equation}
|\tilde{m}\rangle_{\bf L}=a_m\sum_{j=1}^N
\cos\{{\frac{\pi}{2N}(m-1)(2j-1)}\}|{\bf j}\rangle,
\end{equation}
where $m=1,2,...,N$, $a_1=1/\sqrt{N}$ and $a_{m\neq 1}=\sqrt{2/N}$
with energy (on setting $E_0=0$) given by
$E_m=2B+2J(1-\cos\{{\frac{\pi}{N}(m-1)}\})$. In this case,
$f_{r,s}(t_0)$ is given by
\begin{equation}
f_{r,s}(t_0) = \sum_{m=1}^N \langle {\bf r}|\tilde{m}\rangle \langle
\tilde{m}|{\bf s}\rangle e^{-iE_mt_0} = IDCT_s(v_{m,r}) \label{dct}
\end{equation}
where, $v_{m,r}=a_m\cos{\{\frac{\pi}{2N}(m-1)(2r-1)\}}e^{-iE_mt_0}$
and $IDCT_s(v_{m,r})=\sum_{m=1}^N a_m v_{m,r}
\cos{\{\frac{\pi}{2N}(m-1)(2s-1)\}}$ is the $s$th element of the
inverse discrete cosine transform of the vector $\{v_{m,r}\}$.

\begin{figure}
\begin{center}
\includegraphics[width=3in, clip]{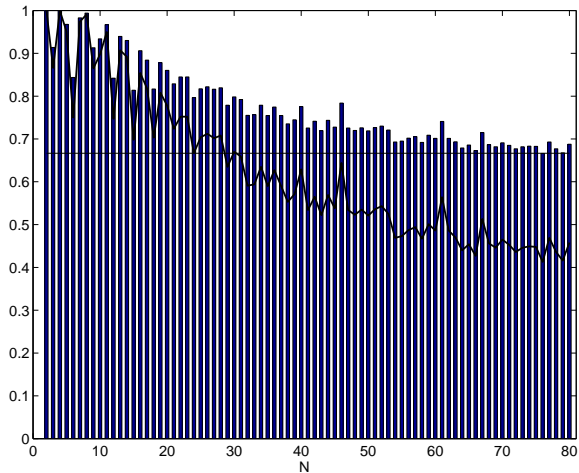}
 \caption{The bar plot shows the maximum
fidelity $F$ of quantum communication and the curve shows the
maximum sharable entanglement ${\cal E}$ achieved in a time interval
$[0,4000/J]$ as a function of the chain length $N$ from $2$ to $80$.
The time $t_0$ at which this maxima is achieved varies with $N$. The
straight line at $F=2/3$ shows the highest fidelity for classical
transmission of a quantum state.}
\label{QT3}
\end{center}
\end{figure}

 We now want to study the performance of our protocol for various
chain lengths $N$ with $s=1$ and $r=N$ (Alice and Bob at opposite
ends of the chain as shown in Fig.\ref{simplest-scheme}). In
general, Bob has to wait for different lengths of time $t_0$ for
different chain lengths $N$, in order to obtain a high fidelity of
quantum state transfer. The time $t_0$ for the highest fidelity in a
given interval of time need not be related to $N$ by a simple
formula. Why is this so? To understand the process, view
Fig.\ref{spinwaves}. As shown in the figure, with an amplitude of
$\cos{\frac{\theta}{2}}$ the spin chain remains in its ground state,
while with an amplitude $e^{i\phi}\sin{\frac{\theta}{2}}$ the spin
chain goes to a time evolving state in which the initial flip at
Alice's end gets spread out (dispersed) and propagates as a series
of waves, some propagating towards and some away from Bob. There is
an initial high amplitude wave-packet which reaches Bob at a time of
about $N/J$, but the highest peaks of $|f_{r,s}|$ at Bob's side are
obtained by a constructive interference of several wave-packets. The
specific time $t_0$ at which such a constructive interference takes
place is not necessarily given by a closed form formula.

    Using Eqs.(\ref{fid}) and (\ref{dct}), we can numerically evaluate
the maximum of $|f_{N,1}(t_0)|$ (which corresponds to the maxima of
both fidelity and entanglement) for various chain lengths from $N=2$
to $N=80$ when Bob is allowed to choose $t_0$ within a finite (but
long) time interval of length $T_{max}=4000/J$. This evaluation is
fast because Eq.(\ref{dct}) allows us to use numerical packages for
the discrete cosine transform. Taking a finite $T_{max}$ is
physically reasonable, as Bob cannot afford to wait indefinitely. It
is to be understood that within $[0,T_{max}]$, the time $t_0$ at
which optimal quantum communication occurs varies with $N$. The
maximum fidelities  as a function of $N$  and the maximum amounts of
entanglement sharable (both rounded to $3$ decimal places) are shown
in Fig.\ref{QT3}.

Fig.\ref{QT3}, shows various interesting features of our protocol.
The plot also shows that in addition to the trivial case of $N=2$,
$N=4$ gives perfect ($F=1.000$) quantum state transfer to $3$
decimal places and $N=8$ gives near perfect ($F=0.994$). The
fidelity also exceeds $0.9$ for $N=7,10,11,13$ and $14$. Till $N=21$
we observe that the fidelities are lower when $N$ is divisible by
$3$ in comparison to the fidelities for $N+1$ and $N+2$. While we do
not have a clear cut explanation of this effect, it is obviously a
link between number theory and constructive interference in a line.
The plot also shows that a chain of $N$ as high as $80$ exceeds the
highest fidelity for classical transmission of the state i.e., $2/3$
in the time interval probed by us.

    As an alternate system, one can also consider a ring of $2N$ spins with
Alice and Bob accessing the spins at diametrically opposite sites
($s=1$, $r=N+1$). In that case, as discussed in Ref.\cite{bose02} we
find that the global maxima of $F$ coincides with that of the open
chain of $N$ qubits described above. This means that by using a ring
one can communicate as efficiently over a distance $r-s=N$ as you
can with a open ended line over a distance $r-s=N-1$. An immediate
implication is that a $4$ spin ring allows perfect communication
between diametrically opposite sites (because a $N=2$ spin chain
does so).

\subsection{Is an arbitrarily long spin chain a true quantum channel?-- An answer through entanglement distribution}
\label{ent-dis}
 With the results of the above section, it is not
conclusive that arbitrarily long spin chains are truly ``quantum"
channels. Especially if even for a long interval $T_{max}$, a 80
spin chain cannot transfer quantum states with a fidelity better
than $2/3$, then can we regard a chain of say a 1000 spins as a
quantum channel? It is true that if we increase $T_{max}$, then
longer and longer chains may yield a fidelity better than $2/3$ due
to constructive interference at some time. But because the time is
being obtained by an optimization over an arbitrarily large
interval, this cannot be either proved or disproved easily.

 We will thus consider a different figure of merit for judging the
performance of the spin chain as a quantum communication channel,
namely the amount of entanglement that can be transmitted through
it. Entanglement is the truly ``quantum" correlations that exist
between two systems when they are in a inseparable state such as
$|\psi^{+}\rangle=\frac{1}{\sqrt 2}(|0 1\rangle+|1 0\rangle)$. The
study of entanglement is a huge area of quantum information science,
and we refer the reader to a review such as
Ref.\cite{Plenio-Vedral-CP}. For our current purposes, it is
sufficient to note that if Alice and Bob held one member each of a
pair of particles in the state $|\psi^{+}\rangle$, then one of them
could transmit a quantum state perfectly to the other using a
celebrated protocol called quantum teleportation \cite{bennett93}
(this transmission also requires some additional classical
communication between Alice and Bob).

\begin{figure}
\begin{center}
\includegraphics[width=3in, clip]{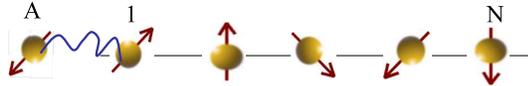}
 \caption{The mechanism of transferring entanglement down a spin chain. The state of one member of
 a pair of qubits in a maximally entangled state is placed on the spin at Alice's end of the chain, while the other member is held
 by Alice. After a while, the spin at Bob's end of the chain will be entangled with the qubit held by Alice.}
\label{enttrans}
\end{center}
\end{figure}

In particular, we will look at the transmission of the state of one
member of a pair of particles in the entangled state
$|\psi^{+}\rangle$ through the spin chain channel. This is the usual
procedure for sharing entanglement between separated parties through
any channel. Alice prepares two qubits in the state
$|\psi^{+}\rangle$, holds one of them (say, $A$) in her hand and
places the other on site $1$ of the chain. This procedure of putting
in one member of an entangled state in a spin chain is illustrated
in Fig.\ref{enttrans}. After waiting for an optimal time $t_0$, Bob
picks up the qubit $N$ from the chain. The joint state shared by
Alice's qubit $A$ and Bob's qubit $N$ at this time is given by
\begin{eqnarray}
\rho_{1N}(t_0)&=&\frac{1}{2}\{(1-|f_{r,s}(t_0)|^2)|00\rangle\langle00|_{1N}\nonumber\\
&+&(|10\rangle+|f_{r,s}(t_0)||01\rangle)(\langle
10|+|f_{r,s}(t_0)|\langle 01|)_{1N}\}\nonumber
\end{eqnarray}
The entanglement ${\cal E}$ of the above state, as quantified by a
certain measure called concurrence \cite{wootters} is given by
\begin{equation}
{\cal E}=|f_{r,s}(t_0)|. \label{ent}
\end{equation}
Thus, for any non-zero $f_{r,s}(t_0)$ (however small), some
entanglement can be shared through the channel. This entanglement,
being that of a $2\times 2$ system, can also be {\em distilled}
\cite{horodecki}. Distillation is a procedure whereby, if several
copies of an entangled state such as $\rho_{1N}(t_0)$ are shared in
parallel between Alice and Bob, then it can be converted to a
smaller collection of pure maximally entangled states
$|\psi^{+}\rangle$ shared between Alice and Bob. This procedure
requires only local actions by Alice and Bob and classical
communication between them. These $|\psi^{+}\rangle$ can
subsequently be used for perfect transmission of a state from Alice
to Bob by quantum teleportation.

  We will now like to estimate the entanglement sharable through chains so large that the optimal $t_0$ identified
  by a numerical search in a long interval of time, is not good enough for an average fidelity higher than $2/3$. We will choose
$t_0$ according to a fixed (in general, non-optimal) prescription
and find how much entanglement can be transferred through the chain.
To motivate this choice, we expand $e^{-iE_mt_0}$ in Eq.(\ref{dct})
as a Bessel function series to obtain
\begin{equation}
{\cal E}=|\sum_{k=-\infty}^\infty (-1)^{Nk}
(J_{N+2Nk}(\beta_0)+iJ^{'}_{N+2Nk}(\beta_0))|, \label{line}
\end{equation}
where $\beta_0=2Jt_0$. Using, $J_N(N+\xi
N^{1/3})\approx(2/N)^{1/3}Ai(-2^{1/3}\xi)$ for large $N$
\cite{childs}, (where $Ai(.)$ is the Airy function) we can prove
that we get a maxima of $J_N(\beta_0)$ at $t_0=(N+0.8089
N^{1/3})/2J$ and at this time
\begin{equation}
{\cal E}\approx 2|J_N(\beta_0)|\approx 1.3499 N^{-1/3},
\end{equation}
which ranges from $0.135$ for $N=1000$ to $1.35\times10^{-4}$ for
$N=10^{12}$ (just $3$ orders decrease in ${\cal E}$ for an increase
in length $N$ by $9$ orders -- a very efficient way to distribute
entanglement). Thus for {\em any} finite $N$, however large, the
chain allows us to distribute entanglement of the order $N^{-1/3}$
in a time $t_0$ linear in $N$. An exchange coupled chain of any
finite length is thus a bonafide quantum channel. It is worthwhile
to point out that for an open ended $XY$ spin chain (Hamiltonian
$H^{XY}$) the amount of entanglement ${\cal E}$ that can be
transmitted by putting one member of an entangled pair of particles
in the spin chain and waiting for a time $t_0=(N+0.8089 N^{1/3})/2J$
is precisely twice the above amount ({\em i.e.,} $2\times1.3499
N^{-1/3}$) \cite{yung03}.

\section{Routes to perfection and near perfection}
The protocols of state transfer and entanglement distribution
described above were primarily motivated by their simplicity. For
the purposes of checking the quality of a qubit array and studying
the quantum response to a quantum impulse -- which are both
important aims, these protocols should suffice. These transfer
schemes, although imperfect, also already motivate the field of
investigating quantum capacities of such spin chain channels
\cite{vittorio04,vittorio-fazio06}. However, for connecting up two
solid state quantum registers, which by far was our main motivation,
it is only perfect or nearly perfect state transfer which is
relevant. For example, to do a quantum gate between qubits in two
separate registers, one needs to transfer the state of a qubit from
the first register to the second perfectly, do a gate between that
qubit and a qubit of the second register, and then transfer its
state back to the first register perfectly. The protocols described
above will not accomplish these tasks on their own. The second
protocol (the one where one transmits the state of one member of a
pair of entangled qubits through the channel) can be made useful
when appended with entanglement distillation procedures. Basically,
one has to first use the spin chain channel repeatedly to make Alice
and Bob share several copies of the partially entangled state
$\rho_{1N}(t_0)$. Then entanglement distillation is performed to
obtain a smaller number of pairs of particles in the state
$|\psi^{+}\rangle$ shared between Alice and Bob. The particles in
the state $|\psi^{+}\rangle$ can now be used to transmit a state
perfectly, do quantum gates between qubits in well separated
locations, and so forth. Clearly, it would be better if we could do
perfect quantum state transfer through a spin chain without invoking
an additional entanglement distillation process. With this view in
mind, several schemes have been proposed which we present below.

\subsection{Engineered chains} The couplings $J_{ij}$ can be
carefully chosen (even when there are only nearest neighbor
couplings, {\em i.e.}, $i=j\pm 1$) to obtain a spin chain which
accomplishes perfect quantum state transfer
\cite{christandl04,petrosyan,albanese04,christandl05}. We need to
briefly digress to one of the very first systems studied in
introductory quantum mechanics, namely a particle in a box, to
clearly understand why this is possible. Let the box be a symmetric
square well centered at $x=0$ ({\em i.e.,} the potential satisfies
$V(-x)=V(x)$) with infinitely high walls. Under these circumstances,
the energy eigenstates $\phi_k(x)$ of $H$ with $k=0,1,...\infty$
satisfy $\phi_k(-x)=(-1)^k\phi_k(x)$ (this property can be called
{\em alternating parity}). The time evolution of {\em any} state
$\psi(x)=\sum_k c_k \phi_k(x)$ is given by
\begin{equation}
\psi(x,t)=\sum_k c_k e^{-iE_k t}\phi_k(x),
\end{equation}
where energies $E_k\propto k^2$ (such energies, which are
proportional to integers, can be called {\em commensurate
energies}). If one chooses a special time of evolution $t=\tau$ such
that $E_k\tau=k^2\pi$, then we have
\begin{equation}
\psi(x,\tau)=\sum_k c_k (-1)^k\phi_k(x)=\sum_k c_k
\phi_k(-x)=\psi(-x).
\end{equation}
Thus there is a time $\tau$ in which the complete wave-function {\em
mirror inverts} about $x=0$. This form of mirror inversion of a
wave-function in an infinite square well have been known for some
time in the quantum optics literature \cite{aronstein97}. The
properties which have been responsible for the mirror inversion
detailed above are the alternating parity and the commensurate
energy properties, and any other Hamiltonian which has the same
features will also exhibit the above mirror inversion.

 There is a close analogy between the position of a particle in a line and a single spin flip in a background of aligned spins
 in a spin chain. The position $x$ of the particle is analogous to the discretized positions $j=1,..,N$ of the
 flip and the wavefunction $\psi(x)$ is simply analogous to a superposition $\sum_jc_j|{\bf j}\rangle$. If the couplings are
   (a) mirror symmetric about the centre of the spin chain ({\em i.e.,} $J_{j,j+1}=J_{N-j,N-j+1}$), which
   gives the alternating parity property, and engineered to (b) give rise to
   commensurate energies, then at a certain time $\tau_c$ the state
   will mirror invert by the logic of the last paragraph, {\em i.e.,} become $\sum_jc_j|{\bf
   N-j}\rangle$. One form of engineering, which works for an open ended $XY$
   spin chain (Hamiltonian $=\sum_j{\bf H}^{XY}_{j,j+1}$) with nearest neighbour interactions
   is $J_{j,j+1}=\sqrt{j(N-j)}$, which gives $E_k\propto k$ \cite{christandl04,petrosyan}. The mirror inversion implies
   that in a time $\tau_c$ a spin flip at $1$st site of the spin chain will be perfectly transferred
   to the $N$th site {\em i.e.,} $f_{1N}(\tau_c)=1$, which in turn
   implies perfect quantum communication and entanglement transfer through formulae of Eqs.(\ref{fid}) and
   (\ref{ent}). Note that engineering a spin chain to obtain a
   commensurate spectrum can give different values of $J_{j,j+1}$
   depending on the specific commensurate spectrum one chooses (such
   as one can choose $E_k\propto k, k^2, k(k+1)$ etc.) and engineered couplings other than
   that mentioned above have also been found \cite{albanese04,karbach,kay06}. In general, one chooses
   a specific commensurate spectrum and then finds the corresponding couplings
   by solving an inverse eigenvalue problem
   \cite{yung05,shi04}. Thus open ended XY spin chain
   with nearly uniform couplings \cite{karbach}, and non-nearest neighbour (dipolar)
   couplings
   \cite{kay06}, have been designed which can all accomplish perfect or near perfect
   quantum state transfer.

\subsection{Wave-packet encoding}
    So far, we have been considering various protocols in which the
      $|1\rangle$ state to be transmitted by the chain is encoded on it as a single spin
    flip at one end. Remembering the analogy between positions $x$ of a particle in a line
     and discretized positions $j$ of a spin flip in a background of aligned spins, this is analogous to an infinitely narrow wavefunction (a Dirac delta function) of a particle in a line.
    For typical potentials such as a particle in a box or a harmonic oscillator, such a wavefunction is known to disperse (spread) rapidly in
    space and thus the particle's behaviour becomes very unlike that of a classical particle. However, it is possible to place particles
    in gaussian wave-packet states which have a low dispersion and travel with a definite group velocity, more or less as a classical
    particle would. Then any information encoded as a superposition of the presence and absence of a particle in such a state would
    travel more or less with a well defined velocity and reach the receiver at a predetermined
    time. Can one try a similar trick in the transmission of quantum
    information down a spin chain? Such a protocol was first
    proposed for a ring of $N$ spins interacting through the Heisenberg or the XY model \cite{osborne04},
    where ``truncated" gaussian wave-packet states $|G(j_0,k)\rangle = \sum_{j}e^{-(j-j_0)^2/L^2}e^{-ik_0j}|{\bf
    j}\rangle$ centred at the site $j=j_0$ (and defined over $L$ sites around the site $j_0$) and with velocity $\propto k_0$ were used.
    In a block of $L$ spins near a site $j_A$, Alice encodes
    the $|1\rangle$ state of the qubit she wants to transmit on
    $|G(j_A,k)\rangle$ (the $|0\rangle$ is encoded as in the
    previous protocols). For appropriate choice of $k_0$ (see
    Ref.\cite{osborne04}) one can choose $L\sim N^{1/3}$, so that this wave-packet travels with
    a dispersion which is negligible and remains constant no matter how
    far it travels down the ring. Thus, Bob located at any distant
    site along the ring can catch almost the entire wave-packet by
    using a sufficiently long block of spins ($\sim N^{1/3}$) to receive the state.
    To obtain the the fidelity of quantum communication
    and entanglement transfer using this wave-packet scheme, one simply has to replace $f_{1N}$
    in the earlier formulae Eqs.(\ref{fid}) and
   (\ref{ent}) by the amplitude of $|G(j_A,k)\rangle$ evolving to
    $|G(j_B,k)\rangle$ with time, where $j_B$ is the site around
    which Bob has access to a block of $\sim N^{1/3}$ spins. This
    amplitude can be as $95$ percent for $N$ up to $5000$
    \cite{osborne04}. For open ended spin chains, it is not ideal for Alice to encode
    $|1\rangle$ in a truncated gaussian wave-packet at one end of
    the chain, as this will distort and spread when it reaches Bob at the other
    end of the chain. This problem has been cleverly resolved in Ref.\cite{haselgrove04}, which shows that
    Alice can encode a different wave-form on a block of $N$ spins at
    one end, which {\em evolves} to a gaussian wave-packet of certain $k_0$ in the
    chain and travels to the other end with minimal dispersion.
    Additionally, Ref.\cite{haselgrove04} points out how such an
    encoding in blocks for near perfect communications is possible for {\em any} graph of
    spins (albeit, connected with each other through interactions) and {\em how} such an encoding/decoding can be accomplished
    by Alice/Bob with access only to a single spin each, but continuous time control of
    the interactions of these spins with the graph. Gaussian
    wave-packet encodings have also been suggested for communication through spin-chains in
    various static external fields \cite{shi04,zong-chen}.

\subsection{Coupling qubits weakly to a quantum many body system}
Another approach is to couple the sending and receiving qubits
weakly to a quantum many body system
\cite{shi04,plenio04,wojcik05,wojcik06}. Say the many body system is
an arbitrary graph of spins which interact with each other with a
coupling strength $J\sim 1$, while the sending and receiving qubits
are coupled to the system through a coupling $\epsilon$ where
$\epsilon<<1$. Moreover, assume all the couplings to be of $XY$ or
Heisenberg type (other interactions would also do as long as they
can enable the transfer of an excitation through the graph from the
sending to the receiving qubit). Then, one can derive {\em
effective} XY or Heisenberg Hamiltonians between the two qubits when
there are {\em no} eigenstates of the many-body system whose energy
is close to $0$ \cite{shi04,wojcik05,wojcik06}. This is possible,
for example, when the many-body system is in its ground state and
has a finite energy gap $\Delta$ between the ground and the first
excited state (such as a spin ladder \cite{shi04}). Effectively, a
Hamiltonian of the form $\epsilon^2 {\bf H}^{XY}_{sr}$ or
$\epsilon^2 {\bf H}_{sr}$ acts on them, where $s$ and $r$ stand for
the sending and the receiving qubits respectively. The effective
Hamiltonian can be rigorously derived using second order
perturbation theory \cite{shi04,wojcik05,wojcik06}. This case, which
we will call the ``off resonant" case \cite{wojcik06}, enables a
perfect quantum state transfer by the simple fact that the two qubit
Hamiltonians $\epsilon^2 {\bf H}^{XY}_{sr}$ or $\epsilon^2 {\bf
H}_{sr}$ do so in a time $t\sim 1/\epsilon^2$. The other case is
when the many-body system has exactly {\em one} available state
$|\lambda\rangle$ of zero energy of the type in which a single spin
is flipped from the ground state, and beyond that, there is a gap
$\Delta$ to all other states of the single flip type. Then a
``resonant" transfer \cite{wojcik06} through the many body system
takes place with an effective Hamiltonian $\epsilon {\bf
H}^{XY}_{sM}+\epsilon {\bf H}^{XY}_{Mr}$, where
$\sigma^{x}_M,\sigma^{y}_M$ and $\sigma^{z}_M$ are defined for a
{\em delocalized qubit} with $\sigma^{z}_M=+1$ corresponding to the
many body system being in $|\lambda\rangle$ and $\sigma^{z}_M=-1$
corresponding to the many body system being in its ground state
\cite{plenio04,wojcik06}. This resonant effective Hamiltonian is
just a three qubit $XY$ spin chain which can perfectly transfer
states in a time scale $t\sim 1/\epsilon$
\cite{christandl04,yung05}.

\subsection{Ising chain with global pulses}
Another approach is to use a spin chain Hamiltonian with a different
type of coupling, namely a nearest neighbour Ising coupling, as
given by
\begin{equation}
{\bf H}^{Ising}=\sum_{j=1}^N J\sigma^z_j \sigma^z_{j+1},
\end{equation}
in conjunction with ``global" pulses (pulses that act on each spin
of the chain in exactly the same way) at regular intervals to
perfectly transport a state from one of its ends to the other
\cite{fitzsimons06}. To understand this, we will need two unitary
operations, one called the Hadamard (denoted by $H$) which acts on a
single qubit to change $|0\rangle$ to
$|+\rangle=|0\rangle+|1\rangle$ and  $|1\rangle$ to
$|-\rangle=|0\rangle-|1\rangle$ and the other called the
Controlled-$Z$ or simply $CZ$, which acts on two qubits to change
$|1\rangle|\pm\rangle$ to $|1\rangle|\mp\rangle$, but keep
$|0\rangle|\pm\rangle$ unchanged \cite{NC}. It is shown in
Ref.\cite{fitzsimons06} that an Ising chain evolving on its own for
a time $\pi/4J$ followed by fast (instantaneous) operations on the
chain by global pulses at time $\pi/4J$ (and some extra operations,
also fast, at the very ends of the spin chain), accomplishes the
operation $S$ which is equivalent to a $CZ$ between all adjacent
pairs of spins followed by a $H$ on each spin. The entire time
evolution with interruptions by the instantaneous pulses at regular
intervals is then equivalent to a series of applications of $S$. Now
imagine a $N=4$ spin chain to be initialized in the state
$(\alpha|0\rangle_1+\beta|1\rangle_1)|+\rangle_2|0\rangle_3|+\rangle_4$.
Then successive applications of four $S$ operations accomplishes the
evolution
\begin{eqnarray}
&(\alpha|0\rangle_1+\beta|1\rangle_1)|+\rangle_2|0\rangle_3|+\rangle_4 \nonumber\\
\overset{S}{\rightarrow} &(\alpha|+\rangle_1|0\rangle_2+\beta|-\rangle_1|1\rangle_2)|+\rangle_3|0\rangle_4 \nonumber\\
\overset{S}{\rightarrow} &|0\rangle_1(\alpha|+\rangle_2|0\rangle_3+\beta|-\rangle_2|1\rangle_3)|+\rangle_4 \nonumber\\
\overset{S}{\rightarrow} &|+\rangle_1|0\rangle_2(\alpha|+\rangle_3|0\rangle_4+\beta|-\rangle_3|1\rangle_4) \nonumber\\
\overset{S}{\rightarrow} &|0\rangle_1|+\rangle_2|0\rangle_3(\alpha|+\rangle_4+\beta|-\rangle_4).\nonumber\\
\end{eqnarray}
Thus an extra $H$ operation on each qubit (accomplishable by global
pulses) after the above evolution has completely transferred the
state at site $1$ to site $4$. The authors of
Ref.\cite{fitzsimons06} show that in general, for a $N$ spin chain,
one is able to transfer a quantum state from one end to the other by
$N$ applications of $S$ ({\em i.e.} evolution of the chain till time
$N\pi/4J$ interrupted by instantaneous pulses at regular intervals)
and a $H$ on each qubit at the end of the evolution. Moreover, they
also show that such a transfer can be accomplished by {\em any}
starting state of the spin chain.

It should also be mentioned here that some very recent works show
that if some degree of slow modulation of the couplings are allowed,
then adiabatic passage can also be used to transfer quantum states
perfectly through a spin chain channel \cite{ohshima07,eckert07}.

\section{Perfection with simplicity: a dual-rail protocol}
One always has to pay a ``price" for perfect transfer. The original
protocol with uniform couplings and single qubit encoding
\cite{bose02} necessitate mixed state entanglement distillation, and
hence several uses of the channel even for the near perfect
transmission of a {\em single} qubit. The engineering of couplings
\cite{christandl04,petrosyan,albanese04,christandl05} will be
naturally restricted to those physical implementations where
interaction strengths can be tuned to appropriate values, as opposed
to being ``given", while using wave-packets
\cite{osborne04,haselgrove04} necessitate involving several qubits
for encoding or continuous time control. Weak couplings
\cite{shi04,plenio04,wojcik05,wojcik06} may give a slower transfer,
while global pulse schemes specialize to Ising chains
\cite{fitzsimons06}. We now present a protocol introduced in
Ref.\cite{burgarth04} which does not have to pay any of the above
prices at the expense of using {\em two} spin chains in parallel as
opposed to a single chain. The couplings in the spin chains need not
be either uniform or specially engineered and could even be mildly
random (with the reasonable assumptions that the chains are {\em
similar} to each other and permit state transfer) \cite{burgarth05}
and the scheme involves only two qubits for encoding. We call this a
dual rail protocol \cite{burgarth04,burgarth05}.

      An assumption of control at either end of the spin chain, of the
spins which Alice and Bob control, is an implicit assumption in all
communication protocols using spin chains. At the very least, Alice
has to ``swap in" a quantum state at one end of the chain (from her
quantum register) which requires a tunable interaction of the first
spin with a register spin. Bob has to have similar ability for
retrieving the qubit from the other end of the chain. The same type
of interaction, namely a switchable two qubit interaction, suffices
for the encoding and decoding of this scheme involving parallel spin
chains. We first describe below how parallel spin chains can be used
for a ``heralded" perfect quantum state transmission, where
conditional on a positive outcome of a measurement, Bob can conclude
that he has accurately received the state transmitted to him.

\begin{figure}
\begin{center}
\rotatebox{-90}{\resizebox{!}{3in}{\includegraphics[width=3in,
clip]{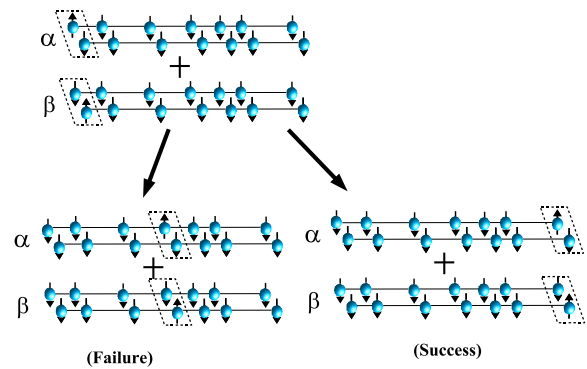}}}
 \caption{The dual rail protocol for perfect quantum communications through spin chain channels. The upper
 part of the figure shows an arbitrary superposition of the logical qubit states being encoded at one end
 of the parallel chains (the dotted box shows the encoded states of a qubit). The lower parts of the figure
 show two possible states of the chain after the passage of some time and Bob's measurement. Bob's success corresponds to
 the state being received perfectly on the spins at his end of the chain. His failure corresponds to the information being retained
 in parts of the chain not accessed by Bob, as shown in the lower left hand side of the figure. Strictly speaking the state corresponding
 to failure will be a superposition of all states of the form shown in the lower left hand side {i.e.,} it will be a superposition of all those states in which
  the
 dotted box is at sites other than $N$.}
\label{parallel}
\end{center}
\end{figure}

\subsection{A heralded perfect state transfer}
The idea of Ref.\cite{burgarth04} is to use two spin chains $I$ and
$II$ in parallel as a single communication channel as shown in
Fig.\ref{parallel}. As in previous protocols, this protocol is also
restricted to a sector in which each spin chain has at most one spin
flipped in a background of spins in the $|0\rangle$ state. We will
thus use a similar notation as before for spin chain states, namely
$|{\bf 0}\rangle^{(I)}$ and $|{\bf 0}\rangle^{(II)}$  denoting the
ferromagnetic ground states (all spins in the $|0\rangle$ state) of
the chains $I$ and $II$ respectively, and $|{\bf j}\rangle^{(I)}$
and $|{\bf j}\rangle^{(II)}$ denoting the $j$th spin flipped to the
$|1\rangle$ state in the chains $I$ and $II$ respectively. For the
moment, we assume the chains to be non-interacting, identical copies
of each other and coupled by uniform Heisenberg or $XY$ interactions
as in the original single chain based protocol \cite{bose02}(some of
these assumptions can be relaxed \cite{burgarth05}, as will be
discussed later). The first spin of each chain is controlled by
Alice, while the $N$th spin of each chain is controlled by Bob.
Initially, the spin chains are assumed to be in the state $|{\bf
0}\rangle^{(I)}$ and $|{\bf 0}\rangle^{(II)}$. When Alice intends to
transmit a qubit state $|\phi\rangle=\alpha |0\rangle +\beta
|1\rangle$, she encodes this into the two spins that she controls as
$|\tilde{\phi}\rangle=\alpha |01\rangle +\beta |10\rangle$. This
encoding can be accomplished by a simple two qubit unitary operation
(a two qubit quantum gate) \cite{burgarth04} involving the qubits
that Alice controls, and can be accomplished inside the quantum
computer. This encoding places the entire system of two spin chains
in the quantum state
\begin{equation}
|\Phi(0)\rangle=\alpha |{\bf 0}\rangle^{(I)}|{\bf 1}\rangle^{(II)}
+\beta |{\bf 1}\rangle^{(I)}|{\bf 0}\rangle^{(II)},
\end{equation}
which evolves with time as
\begin{equation}
|\Phi(t)\rangle=\sum_{j=1}^Nf_{1j}(t)(\alpha |{\bf 0}\rangle^{(I)}
|{\bf j}\rangle^{(II)} +\beta |{\bf j}\rangle^{(I)}|{\bf
0}\rangle^{(II)}). \label{phit}
\end{equation}
It is then simple to calculate, by using the method of partial
tracing as discussed before, the time varying density operator
$\varrho(t)$ of the two spins which Bob controls, and is found to be
\begin{equation}
\varrho(t)=(1-|f_{1N}(t)|^2)|00\rangle\langle
00|+|f_{1N}(t)|^2|\tilde{\phi}\rangle\langle \tilde{\phi}|.
\end{equation}
Bob now measures the ``total" spin component of his spins in the $z$
direction {\em without} measuring any of the spins individually.
Such a measurement gives a value $-1$ for $|00\rangle$ and the value
$0$ for any superposition of $|01\rangle$ and $|10\rangle$.
Physically, such a measurement can be accomplished by a coarse
grained spin measurement on the two spins of Bob which is
insensitive to the precise location of the magnetic moment (a parity
measurement on the two spins \cite{loss}, which gives an outcome $0$
for $|00\rangle$ and an outcome $1$ for any superposition of
$|01\rangle$ and $|10\rangle$ also suffices). When Bob gets the
outcome $-1$, which happens with probability $(1-|f_{1N}(t)|^2)$,
his spins are projected to the state $|00\rangle$ irrespective of
the state transmitted. This heralds the failure on Bob's part to
receive the state. On the other hand, when he obtains the outcome
$0$, which happens with probability $|f_{1N}(t)|^2$, his spins are
projected to the state $|\tilde{\phi}\rangle$. Bob can now simply
apply the inverse of the quantum gate that Alice used for encoding
to his spins to obtain a decoded state $|\phi\rangle$, which
corresponds to his perfect retrieval the state transmitted by Alice.
Here we should point out that Bob's actions could also be combined
into one two qubit quantum gate followed by a single qubit
measurement, which the reader will find in the original presentation
of this protocol \cite{burgarth04}. For long Heisenberg and $XY$
chains, thus, Bob's success probability in this heralded scheme
scales as $|f_{1N}(t)|^2\sim 1/N^{2/3}$ in a time $t\sim O(N/J)$. In
some sense, we have been able to {\em convert} the partial fidelity
of the transmitted state in the original spin chain communication
scheme to a probability of success, and when successful, Bob
receives the transmitted state perfectly. It is easy to verify that
all the above also holds for transmitting entanglement through the
parallel spin chain channel. Thus if Alice and Bob were merely
attempting to establish entanglement, then Alice could try to send
the state of one member of an entangled pair of qubits through the
channel. If Bob fails to receive the state, the channel is reset to
the state $|{\bf 0}\rangle^{(I)}|{\bf 0}\rangle^{(II)}$ (by cooling
to the ground state in a magnetic field, for example) and Alice
tries to transmit entanglement again. On average after attempting
about $N^{2/3}$ times, each of which takes about $t\sim O(N/J)$
amount of time (so that the total time is $O(N^{1.67}/J)$), Alice
and Bob will be able to share a pure maximally entangled state of
two qubits such as $|\psi^{+}\rangle$.
\subsection{Unlimited enhancement of success probability by waiting}
Interestingly, if one was willing to wait till a time
$O(N^{1.67}/J)$, then even the process of resetting the channel and
repeatedly attempting to transmit the state (or entanglement, as
described above) becomes unnecessary. When Bob fails, the state of
the parallel spin chains is $\sum_{j=1}^{N-1}f_{1j}(t)(\alpha |{\bf
0}\rangle^{(I)} |{\bf j}\rangle^{(II)} +\beta |{\bf
j}\rangle^{(I)}|{\bf 0}\rangle^{(II)})$. If we relabel
$\sum_{j=1}^{N-1} f_{1j}|{\bf j}\rangle^{(I)}$ and $\sum_{j=1}^{N-1}
f_{1j}|{\bf j}\rangle^{(II)}$ as $|\varphi(t)\rangle^{(I)}$ and
$|\varphi(t)\rangle^{(II)}$, then this state can be rewritten as
$\alpha |{\bf 0}\rangle^{(I)} |\varphi(t)\rangle^{(II)} +\beta
|\varphi(t)\rangle^{(I)}|{\bf 0}\rangle^{(II)}$, which immediately
clarifies to the reader that the initial quantum information is
unspoilt and simply encoded in a delocalized form in the two chains.
So can Bob try to retrieve the state again after waiting for a
while? The state of the parallel chain system evolves, in another
time $\tau$, to a state of the form
$\sum_{j=1}^{N}\tilde{f_{1j}}(\alpha |{\bf 0}\rangle^{(I)}|{\bf
j}\rangle^{(II)} +\beta |{\bf j}\rangle^{(I)}|{\bf
0}\rangle^{(II)})$, where $\tilde{f_{1j}}$ can be simply expressed
in terms of $f_{ij}(t)$ and $f_{ij}(\tau)$ \cite{burgarth04}. This
is just Eq.(\ref{phit}), with $f_{1j}(t)$ replaced by
$\tilde{f_{1j}}$. Thus, Bob's actions (measurement and decoding) may
again be repeated at a time $\tau$ after a failure, and again there
is a probability of success equal to $|\tilde{f_{1N}}|^2$. In this
way, whenever Bob fails, he simply waits and again attempts to
retrieve the state from the chain. In Ref.\cite{burgarth04} it has
been argued that the total probablity of success on repeated
measurements by Bob can be made as high as $0.99$ in a time scale of
$\sim O(N^{1.67}/J)$.

   Having read through the dual rail protocol, the reader may now ask some
natural questions. For example, what happens if measurements of the
same nature as Bob's were carried out at regular intervals even at
other sites of the spin chain. Preliminary results indicate that the
state transfer may then be significantly speeded up \cite{vaucher}.
Another interesting question is whether there is any gain in going
to multiple parallel chains as opposed to two? Indeed there is a
gain in efficiency. While the dual rail protocol uses two spin chain
channels to transmit a single
  qubit, multiple rails can be used to transmit a qubit per
  chain for a large number of rails if the states of multiple qubits are suitably encoded on Alice's end
  of the chain \cite{vittorio05}. Recently, it has also been shown that even if the instantaneous measurements at fixed instants
  of time in the dual rail protocol were replaced by more realistic finite strength continuous measurements at the receiving end, the performance
  of the protocol can remain similar as long as of the strength of the measurement is appropriately tuned
  \cite{jacobs07}.

\subsection{Inherent robust aspects of the dual rail protocol}
The dual rail protocol is intrinsically robust in many ways. Suppose
Bob has not measured his spins precisely at the optimal time, at
which his probability of success is highest, but slightly before or
after that time. The probability will still be quite high (as it is
an analytic function of time), and when successful, Bob will still
receive an unspoilt version of the state transmitted by Alice. This
contrasts all previous protocols in which the fidelity itself of the
transmitted state is affected by the time of its reception, and a
non-optimal time results in the state being received by Bob being a
somewhat corrupted version of the state transmitted by Alice.
Another important robust aspect of the protocol stems from the fact
the quantum state is transmitted through the parallel chains through
what is called a ``decoherence free" encoding. If the parallel
chains are not that distant (this may be needed anyway to ensure the
possibility of the quantum gates or joint measurements at the ends),
any external environment couples to them only through their {\em
net} magnetic moment in some given direction, such as through the
operator $\sigma_z^{(I)}+\sigma_z^{(II)}$. Alice encodes the state
to be transmitted on her two spins as a superposition of states
$|01\rangle$ and $|10\rangle$, which are eigenstates of
$\sigma_z^{(I)}+\sigma_z^{(II)}$ with eigenvalue $0$, and hence
decoupled from the environment. The same holds {\em during} the
transmission, as superpositions of states $|{\bf 0}\rangle^{(I)}
|{\bf j}\rangle^{(II)}$ and $|{\bf j}\rangle^{(I)}|{\bf
0}\rangle^{(II)}$ are similarly decoupled from the environment. This
decoupling will enhance the time-scale over which the behavior of
the parallel chain scheme is unaffected by an external environment
relative to the time scale over which single spin chain based
communication schemes remain unaffected. The above robustness
aspects have been pointed out in more detail in
Ref.\cite{burgarth04}, while Ref.\cite{burgarth05} points out that
the schemes are robust even to a mismatch of the chains with each
other (the chains do not need to be identical copies of each other).
In a nutshell, for mismatched rails which are not too dissimilar,
there will be a series of times at which the absolute values of the
amplitudes $f_{1N}^{(I)}$ and $f_{1N}^{(II)}$ for the transmission
of a flip from site $1$ to $N$ through the chains $I$ and $II$ will
be coincident. If Bob performs his actions to receive the state at
these specific times, then conditional on success, he will still
receive an uncorrupted state \cite{burgarth05}.

\section{Single chains for perfect transfer}
It is possible to obtain some protocols with similar positive
attributes as the dual rail protocol which use only a single spin
chain, which we discuss below.

\subsection{Chain of coupled qutrits}
\label{perm} Firstly, it is possible to use a single chain of higher
dimensional quantum systems, such as qutrits (quantum three level
systems) with levels $|+1\rangle,|0\rangle$ and $|-1\rangle$,
instead of two parallel chains \cite{burgarth04}. For our protocol,
the qutrits should be coupled by the natural generalization of an
exchange (or isotropic Heisenberg) interaction to higher dimensions
given by a Hamiltonian $H=\sum_{i}P_{i,i+1}$ where
\begin{equation}
P_{i,j}|\psi\rangle_i|\phi\rangle_j=|\phi\rangle_i|\psi\rangle_j.
\end{equation}
One can look up Ref.\cite{hadley05} and references therein for a
more detailed discussion of the above Hamiltonian. Please note
carefully that in general, the above is {\em not} a Hamiltonian of a
chain of coupled spin-1 systems (except for very special cases), but
there are physical systems such as optical lattices, where it may be
found \cite{hadley06}. For such a Hamiltonian, the state $|{\bf
0}\rangle$ in which each qutrit is in the $|0\rangle$ state, is an
eigenstate. From this state, one generates the states $|{\bf
+j}\rangle$ and $|{\bf -j}\rangle$ of the chain in which the $j$th
qutrit is flipped to the $|+1\rangle$ and $|-1\rangle$ state
respectively. Then the dual rail protocol described in the previous
section can be exactly adapted to the chain of qutrits with the
mappings $|{\bf 0}\rangle^{(I)} |{\bf
0}\rangle^{(II)}\rightarrow|{\bf 0}\rangle$,$|{\bf 0}\rangle^{(I)}
|{\bf j}\rangle^{(II)}\rightarrow|{\bf +j}\rangle$ and $|{\bf
j}\rangle^{(I)} |{\bf 0}\rangle^{(II)}\rightarrow|{\bf -j}\rangle$.
One can check that Bob's measurement will now be mapped to a
measurement which finds out whether his qutrit is in the state
$|0\rangle$ or not (without ascertaining whether the qutrit is in
the state $|+1\rangle$ or $|-1\rangle$) and success is when he
obtains the result ``not $|0\rangle$". In a similar manner, if one
had exchange coupled $d+1$ level systems, one could use one of those
levels as the $|0\rangle$ state, and use the others to transmit a
$d$ dimensional system perfectly with $0.99$ probability of success
in a time $\sim O(N^{1.67}/J)$ through the chain.
\subsection{Receiver with memory}
Curiously enough, even a single spin-1/2 chain without any encoding
from Alice, can transmit a quantum state perfectly to Bob if he had
a memory at his disposal \cite{burgarth-giovannetti}. As in all
unencoded single chain protocols describe before, Alice simply
places a state $\alpha |0\rangle +\beta |1\rangle$ on one end of a
spin chain initialized in the state $|{\bf 0}\rangle$. The clever
trick used for receiving the state with arbitrarily high fidelity is
to swap the state of Bob's qubit with that of a ``fresh" memory
qubit in state $|0\rangle$ at regular intervals
\cite{burgarth-giovannetti}. The memory qubits are always
non-interacting with each other and also non-interacting with the
chain apart from during the swaps. Eventually, in a time scale which
has been argued in Ref.\cite{burgarth-giovannetti} to be no larger
than $O(N^2/J)$ for a broad class of chains (not necessarily uniform
or Heisenberg or XY coupled), the spin chain ends up in the state
$|{\bf 0}\rangle$. In other words, all information about the input
state is erased from the chain and transferred entirely to the
collection of memory qubits. The unitarity of the whole evolution
(the spin chain dynamics and the series of unitary swap operations)
then guarantees that the collective state of all memory qubits is a
function of $\alpha |0\rangle +\beta |1\rangle$. The same unitarity
also guarantees that Bob can use another unitary operation to
convert the state from a multiple qubit memory state to the single
qubit state $\alpha |0\rangle +\beta |1\rangle$, thereby completing
the reception of the state transmitted by Alice.
Ref.\cite{burgarth-giovannetti} also shows that Alice can transmit
many qubits simultaneously through the chain using the above
protocol. Very recently there has been another interesting proposal
in which Bob need possess only one memory qubit to receive a single
qubit transmitted by Alice \cite{burgarth-giovannetti-bose}. Bob
lets this memory qubit interact with the spin at his end of the
chain at regular intervals, but for different durations of time
during each interaction. These times durations are so chosen that
the entire amplitude of Bob's spin to be in the $|1\rangle$ state is
transferred to the memory qubit (the reader may check
Ref.\cite{burgarth-giovannetti-bose} to satisfy him/herself that
this is indeed possible through an unitary operation). In this way,
as before, the chain will finally be left in the state $|{\bf
0}\rangle$ (all information erased), and the memory qubit will end
up in the state $\alpha|0\rangle+\beta|1\rangle$. One positive
feature of this scheme is that the memory qubit may itself be a part
of the spin chain, say an extra $N+1$th spin attached to the $N$th
spin of the chain, with its interaction with the chain being
switchable through a local magnetic field
\cite{burgarth-giovannetti-bose}.

\section{Physical Implementations}
There has been several suggestions for the physical implementations
of the quantum communication schemes described in this review.
Essentially, chains of any physical system which has been proposed
as a qubit, and which can be coupled with each other through an
appropriate interaction (such as Heisenberg or XY), can be used.
However, a permanent (non-tunable) coupling between the qubits will
suffice. The most prominent class of suggestions are based on chains
of superconducting qubits \cite{romito,bruder05,bruder06}. One such
example is based on charge qubits \cite{romito}, where the two
states of the qubit are the presence or absence of a Cooper pair in
a superconducting island. A Cooper pair can hop from one island to
its neighbour through a Josephson coupling, which acts as an XY term
in the Hamiltonian. There are additional parts to this Hamiltonian,
such as a long range (much more than nearest neighbour) Coulomb
interaction, which cannot be ignored \cite{romito}. Alternatively,
one can use two opposite flux states of superconducting rings as the
two states of a qubit, while these rings are coupled to each other
through capacitive couplings \cite{bruder05,bruder06}.
Implementations of slightly different schemes for entanglement
distribution have also been discussed for Josephson junction arrays
\cite{huelga}.

 From the point of view of simulations with short chains, NMR is well
suited (the ability to simulate communications through a 6 spin
Heisenberg ring using benzene was already suggested in
Ref.\cite{bose02}). Recently, quantum communication through a
3-qubit Ising chain using global pulses (the scheme of
Ref.\cite{fitzsimons06}) was demonstrated using NMR
\cite{fitzsimons-jones}. In quantum dots, which are tunnel coupled,
so that electrons can freely hop from one dot to another, the
electronic spin may be used as a qubit, and its transport in an
array has been studied in a scheme slightly different from the ones
discussed in this review \cite{petrosyan}. Alternatively, excitons
(coupled electron and hole pairs) in quantum dots may be used for
implementing the schemes described in this review \cite{dAmico} (and
other communication schemes as well \cite{dAmico-brendon}), with the
two states of the qubit being the presence or absence of an exciton
in a quantum dot, and an XY coupling between neighbouring dots being
provided by the F\"{o}rster interaction. There have also been
various suggestions for implementations of the dynamics of XY
  chains in other systems, such as an array of low loss cavities for holding light coupled to
  each other so that photons can freely hop between them. When the light in each cavity is coupled to a single two-level system, then
  the system simulates a XY model \cite{angelakis}, and
  consequently, the schemes described here can also be implemented in such arrays.
Another example is the simulation with chains of atoms trapped in
optical lattices \cite{duan-lukin-demler}, which should again be an
avenue for implementation.

\section{Developing and future directions}
We would like to end the review by briefly pointing out the varied
directions in which the topic is expanding as these offer the scope
of much future work. One of the most obvious questions is what apart
from quantum communications can be accomplished in the same spirit?
By the ``same spirit" we mean through the natural time evolution of
a complex many body system. Starting from the most modest of aims,
one can use a small ring of permanently coupled spins with a flux in
the middle to design a quantum router \cite{korepin04}. In such a
router the communication can be directed between any chosen pair of
users from a multitude of users by adjusting the flux (an implicit
assumption here is that it applies to those spin systems which
involve charged entities at some level, so that a flux ``twists" the
boundary conditions of the ring). Routers have also been proposed in
the context of weakly coupled sending/receiving qubit schemes
\cite{wojcik06}. Permanently coupled rings of spins can serve as
quantum memories \cite{zong-sun}, and a time varying flux through
such rings can undo the natural dispersion of quantum information
stored in individual spins in such a system \cite{illuminati} (in
general, spin rings with a flux is itself emerging as quite a
fruitful system for varied quantum information applications
\cite{korepin04,illuminati,kay05,nori}). Networks of perpetually
coupled spins can also serve as a quantum cloning machine, were
information initially placed on $N$ of the spins is cloned to $M$ of
the spins due to the natural dynamical evolution of the network
\cite{palma04,palma05,du06}. Simple spin networks can also be
automata for single spin measurement \cite{kay2006a}, while certain
other desirable automata have been shown to be impossible
\cite{kay07}. What other dedicated small scale applications can we
find for small (possibly engineered) networks of spins can thus be
an interesting future goal.

   Of course a more ambitious goal is to achieve full-scale quantum computation
   using permanently coupled systems such as spin chains. Indeed, it was noticed quite
   early (even predating the suggestion of spin chains for quantum communications) that
   the free evolution of small segments of $3-5$ Heisenberg/XY interacting spins can give
   rise to quantum gates between qubits encoded in these segments
   \cite{yung03,benjamin-bose}, which was exploited for designing a universal quantum
   computer with spin chains \cite{benjamin-bose}. Recently, natural
   evolutions of designer spin networks for quantum gates where qubits are fed
   in from one end of the system and read out from the other end
   of the system after gates have acted on them, have been proposed
   \cite{kay05}. Interestingly, the free evolution of engineered spin chains can enact
   interesting multi-qubit quantum gates because of the exchange of fermionic operators to which
   such systems can be mapped \cite{yung05,Clark04,yung06}.
   Particularly, such spin chains can be used as a processor core for a quantum computer on to which states of qubits are loaded for running certain important classes of algorithms \cite{yung06}.
   Whether one can find a general purpose
   processor core or whether a single permanently coupled network can be designed for running an entire quantum algorithm involving several gates,
   are interesting open questions.

 Spin chains need not act merely as passive buses for quantum information, they can also act as sources
 of entanglement when put in an appropriate initial state. What we really want is a state which dynamically evolves in time
 and generates significant entanglement between the remotest parts of a spin chain. Probably the simplest example
 is flipping the spin at the middle of a ferromagnetic spin chain, and letting the state evolve, which can entangle
 the spins at the opposite ends of the chain
 \cite{yung05}. Such studies have been conducted from very early on
 in the context of harmonic oscillator chains
 \cite{eisert04prl,eisert04} and also in context of graphs of
 qutrits \cite{hadley05} and oscillators \cite{perales05}, and more recently, also for graphs of spins \cite{dAmico07} (see also Ref.\cite{Ujjwal-Aditi} for entanglement from dynamics). However,
 this area, which one can call ``quantum wires for entanglement generation {\em
 and}
 distribution", is open for future work as the possibilities of
 initial states of spin chains are enormous.

  Another area offering possibilities for further exploration is
  when the spin chains, instead of being completely unmodulated, are
  subjected to a time varying external field. Pulsing the whole chain is an example, which is still
  a minimal procedure in comparison to switching individual interactions on and off. We have already
  encountered the fact that regular pulsing in an Ising chain can
  lead to perfect state transfer \cite{fitzsimons06}. The same
  chain on appropriate pulsing can also accomplish universal quantum
  computation \cite{fitzsimons06} (see also the work in
  Ref.\cite{raussendorf}).
  For Heisenberg chains, pulsing
  with a field of an appropriate profile
  after flipping the spin at the middle of a ferromagnetic spin
  chain, gives rise to oppositely propagating entangled gaussian
  wave-packets, which would be very useful for the distribution of
  entanglement \cite{boness06}. Applying external fields to dimerized spin chains can also enable encoding
  qubits in domain walls and by varying the profile of this field, such qubits can be transported \cite{levy}. It remains an open question as to
  what else can be accomplished by time varying external fields on spin
  chains. This question is particularly interesting because
  modulating a few external parameters can control how correlations propagate
  in spin chains \cite{cubitt07}, and the implications of that for
  quantum communications is worthwhile to examine.

 Another open area is related to the issue of quantum response of a physical system to a quantum impulse
 as mentioned in the introduction. Quantum communication through a
 spin chain is one example of this, with the fidelity of
 transmission being a kind of response function. In this context,
 Ref.\cite{hartmann05} finds, for the scenario where Alice's and
 Bob's qubits are weakly coupled to a spin system, that quantum phase
 transitions of the spin system can be detected by a drop in this response function.
 When Alice's and
 Bob's spins are coupled as strongly to the spin system as the spins in the system are coupled to each
 other, then
 Ref.\cite{sanpera} reports the opposite behaviour, namely that the
 same response function peaks at some quantum phase transitions.
 Ref.\cite{sanpera} is also an example of proceeding to higher spins
 in the context of quantum communication using spin chains. This is
 also an interesting direction. It has been shown, for example, that
 chains of higher dimensional systems coupled through the permutation Hamiltonian of
 subsection \ref{perm} can be used to distribute much more entanglement than possible through a spin-1/2 chain \cite{bayat07}. Another example is a lattice system in which a
 large number of bosons can sit at each site, which has been shown to be able to distribute more and more
 entanglement as the number of bosons is increased \cite{bose06}.
 Whether generically chains of higher spin systems provides a better quantum communication bus in comparison to low spin systems is an interesting open question.

  Some directions of investigation are important for realistic physical implementations. For example, what
  happens when spins interact through long range dipolar
  interactions? Then it has been found that state transfer process in an uniformly coupled chain reaches near unit fidelity at the expense of a time
  which scales as $N^3/J$ \cite{avellino06}. Effect of randomness \cite{dechiara-fractal,keating06} and defects \cite{plastina} on the quality of quantum
  information transmission has also been studied, as well thermal effects \cite{bayat}. Of course, a question
  of central importance is what happens if the spins of the chain are not isolated, but coupled to their environment, as such a coupling might be
  unavoidable in certain physical implementations. When each spin of an
  exchange coupled system is coupled to its own independent bath of
  polarized spins
  through a XY coupling, then, it has been shown that the fidelity
  of
  communication is unaffected \cite{burgarth-bose06}. Only the time of communication is delayed and the
  fidelity undergoes rapid oscillations with time
  \cite{burgarth-bose06}. Effects of other kinds of baths on spin chain quantum communications have also
  been recently analyzed \cite{guo06,sun-dec}. No doubt the
  investigations of the above kinds of issues which will
  automatically arise in the context of practical implementations, will form a major part of future
  research on quantum communications through spin chain dynamics.

  \section{Acknowledgements}
  I would like to particularly thank Daniel Burgarth, the chief architect of the dual rail protocol, and my collaborator in
  many of the papers covered here. I would also like
  to thank my other collaborators in various papers mentioned in this review,
  namely, Vittorio Giovannetti, Vladimir Korepin, Vlatko Vedral, Man-Hong Yung, Simon Benjamin,
  Debbie Leung,
   Christopher Hadley, Alessio Serafini, Dimitris Angelakis, Kurt Jacobs, Kosuke Shizume, Bai-Qi Jin, Yasser Omar, Antonio Costa, Jens Eisert,
  Martin Plenio, Benoit Vaucher, Tania Montiero, Tom Boness, Andrew
  Fisher and Martina Avellino. I thank EPSRC for an Advanced Research Fellowship and for support through the grant
GR/S62796/01 and the QIPIRC (GR/S82176/01).

%\end{multicols}

\end{document}